\documentclass[preprint2,11pt]{aastex}
\usepackage{natbib}\usepackage{txfonts}\usepackage{balance}
\usepackage{graphicx}
\newcommand{\eqb}{\begin{eqnarray}}
\newcommand{\eqe}{\end{eqnarray}}
\newcommand{\rlight}{r_{\rm L}}
\begin{document}
\title{
Observational Constraints on Pulsar Wind Theories
}


\author{
J.G. Kirk  }


\affil{
Max-Planck-Institut f\"ur Kernphysik,
Postfach 10 39 80,
69029 Heidelberg,
Germany} 
\email{john.kirk@mpi-hd.mpg.de}

\begin{abstract}
Two-dimensional, relativistic, MHD simulations of pulsar-wind powered 
nebulae provide
strong constraints on the properties of the winds themselves. In particular,
they confirm that Poynting flux must be converted 
into particle energy close to or inside the termination shock
front, emphasising the puzzle known as the $\sigma$ paradox. 
To distinguish between the different possible resolutions of this paradox, 
additional observational constraints are required. 
In this paper, I briefly discuss 
two recents developments in this respect: 
the modelling of high time-resolution optical polarimetry
of the Crab pulsar, and the detection of the 
pulsar/Be~star
binary PSR~1259-63 in TeV energy gamma-rays.
\end{abstract}

\section{Introduction}
Our view of the nebulae powered by pulsar winds has matured dramatically in
the past few years, largely as a result of high resolution images
in the X-ray band
\citep{weisskopfetal00,gaenslerpivovaroffgarmire01,helfandgotthelfhalpern01,gaensleretal02,luetal02,gaensleretal03,pavlovetal03}
complemented, in the case of the Crab Nebula, by high resolution optical images
\citep{hesteretal95,hesteretal02}.
This has motivated at least three
groups of researchers
to perform 2D, relativistic MHD 
simulations
\citep{komissarovlyubarsky03,komissarovlyubarsky04,delzannaetal04,bogovalovetal05}.
The pulsar wind provides the inner boundary conditions for these simulations,
which are concerned with the subsonic flow downstream of the wind's
termination shock front. 
Contact with observation is established by computing the 
synchrotron emissivity of the flows, assuming injection of relativistic
electrons with a prescribed spectrum at the termination shock.
Overall, the observed synchrotron 
morphology, with its rings, asymmetric torii and jet structures,
is fairly convincingly reproduced,
although some of the details still evade explanation
\citep{shibataetal03,morietal04}, and some interesting phenomena demand a
kinetic approach \citep{spitkovskyarons04}. 
There is a consensus on the  
requirements these simulations 
place on the properties of the pulsar wind: 
the energy
flux must be concentrated in the pulsar's 
equatorial plane, and, at least in the case of the Crab, 
the fraction of this energy
transported as Poynting flux must be small. 
This fraction is measured 
by an average over the 
surface of the shock of the $\sigma$ parameter, defined as
the ratio in the comoving frame of the magnetic enthalpy density $B^2/4\pi$ to
the enthalpy density of the plasma, including its rest-mass. For the Crab
Nebula, observations suggest $\sigma\approx 10^{-3}$.
This is puzzling, because, on the one hand, 
standard theories of pair production by pulsars 
\citep{daughertyharding82,hibschmanarons01a,hibschmanarons01b} imply a large
value of $\sigma$ at the point where the wind is launched, and, on the other, 
because a relativistic, ideal MHD wind does not appear to permit a significant
reduction of $\sigma$ 
during propagation \citep{bogovalov01}. This puzzle is not new
\cite[e.g.,][]{kennelcoroniti84b}, and has been termed the \lq\lq $\sigma$ 
paradox\rq\rq,
or \lq\lq $\sigma$ problem\rq\rq

Several resolutions of the paradox have been proposed. It is possible
that magnetic collimation may indeed result in the desired conversion, if the
initial conditions in the wind are sufficiently 
anisotropic \citep{vlahakis04}. Alternatively, the conversion may take place in
a thin layer associated with the termination shock itself
\citep{lyubarsky03b}, or may even be to some extent hidden in the downstream
(nebular) flow \citep{begelman98,shibataetal03}.

A hypothesis that has recently received attention is that dissipation
within the wind zone causes the conversion. 
\citet{coroniti90} and \citet{michel94}
originally pointed out that this might happen 
at the boundary between layers of magnetic field of opposite polarity in the 
\lq\lq striped wind\rq\rq. However, the associated 
acceleration of the wind causes a 
significant dilation of the dissipation timescale, an
effect omitted in the earlier calculations. 
Assuming a gradual acceleration process, it is possible to 
calculate the
evolution of the entropy wave that constitutes the stripes in  
a short wavelength approximation \citep{lyubarskykirk01}. 
Solutions have been found
\citep{kirkskjaeraasen03} 
in which the Lorentz factor $\Gamma$ 
varies as a power of the radius $r$:
\eqb
\Gamma&\propto& r^{q}\textrm{\ with\ }1/3<q<1/2.
\label{similarity}
\eqe
The dissipation proceeds fastest in the solution with the slowest
acceleration, $\Gamma\propto r^{1/3}$. Complete conversion of magnetic into
particle energy before 
reaching the termination shock was found to be 
marginally possible in the case of the Crab pulsar wind. 

Which, if any, of these possibilities will provide the definitive resolution
of the paradox can only be 
decided with the help of observational constraints. These are difficult to 
find. The wind itself, if mostly cold, 
does not radiate synchrotron radiation and so it is difficult to determine its
structure. Recently, however, there have been two observational 
developments that do
appear to hold out the promise of useful constraints.  

In the first, high time-resolution optical polarimetry of the Crab pulsar 
have become available that find a plausible explanation 
under the
hypothesis that the high energy pulses from the Crab pulsar are formed in the 
pulsar wind. If this interpretation is correct, the high-energy pulsed
radiation from this and other pulsars may provide us with 
a direct view of the conversion process itself. 

The second development concerns the 
detection in TeV gamma-rays of the binary system PSR~1259-63. This offers
a diagnostic of the pulsar wind as it enters its termination shock.
Because the
stand-off distance of the shock from the pulsar 
varies with binary phase, it may prove possible to extract the dependence of
the wind's Lorentz factor on distance from the pulsar, thus testing the
acceleration models.

\begin{figure}[t!]
\includegraphics[width=0.5\textwidth,clip=true]{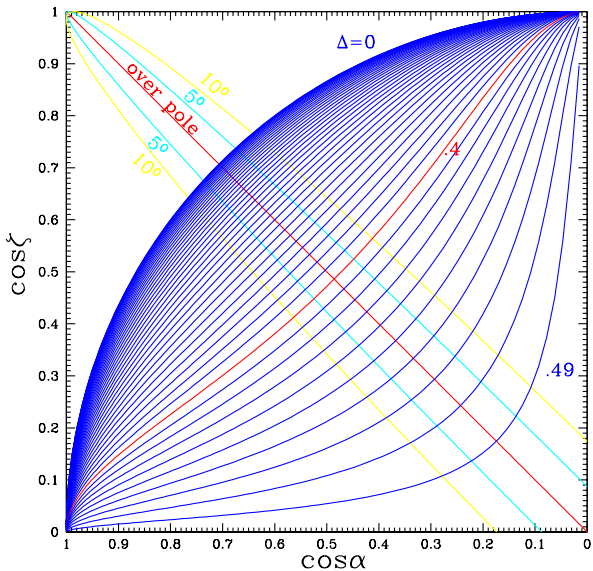}%
\caption{\footnotesize%
\label{inclinations}%
The parameter space of pulsar 
inclination angles $\zeta$ (between the rotation axis and the line of sight)
and
$\alpha$ (between the rotation axis and the magnetic axis). Lines joining the
lower left with the upper right corners (blue and red on-line) 
are contours of constant phase separation $\Delta$
of the twin pulses produced in the striped wind model 
(equally spaced by $0.01$ 
between 
$\Delta=0$ and $\Delta=0.49$).
Lines joining the upper left with the lower right corners are lines of equal 
\lq\lq impact parameter\rq\rq, i.e., minimum angle between the line of sight
and the magnetic axis. The diagonal (red on-line) corresponds to a line of
sight that passes directly over the magnetic pole, the other two pairs to 
minimum angles of $5\degr$ (cyan on-line) and $10\degr$ (yellow on-line).   
The spacing of the Crab pulses, $0.4$, corresponds to the labelled contour
(red on-line) and the $\zeta\approx60\degr$ and $\alpha\approx60\degr$ 
angles favoured by the 
emission model occur at the intersection of this contour with the 
\lq\lq over pole\rq\rq\ diagonal. 
}
\end{figure}
\section{Polarisation of optical pulses from the Crab pulsar}
  
If the hot particles between the stripes of a relativistic, 
radial wind emit radiation, 
the effect of Doppler boosting will cause it to appear
to the observer to be pulsed, provided that the radius $r$ of the radiating
zone satisfies
\eqb
{r\over \rlight}&<& \Gamma^2
\eqe
where $\rlight$ is the radius at 
which the corotation speed would become luminal and $\Gamma$ is the 
bulk Lorentz
factor of the wind.
This led to the suggestion
\citep{kirkskjaeraasengallant02} that the twin-pulse profile
observed in the Crab pulsar, as well as
in the gamma-ray emission of other pulsars, is a manifestation
of the fact that 
two sections of the current sheet are visible per period  
in the striped wind model.
If this interpretation is correct, it constrains the 
the angles $\zeta$ between the rotation
axis of the pulsar and the line of sight and $\alpha$,
between the magnetic and rotation axes, as illustrated in 
Fig.~\ref{inclinations}.

\begin{figure}[t!]
\includegraphics[width=0.5\textwidth,clip=true]{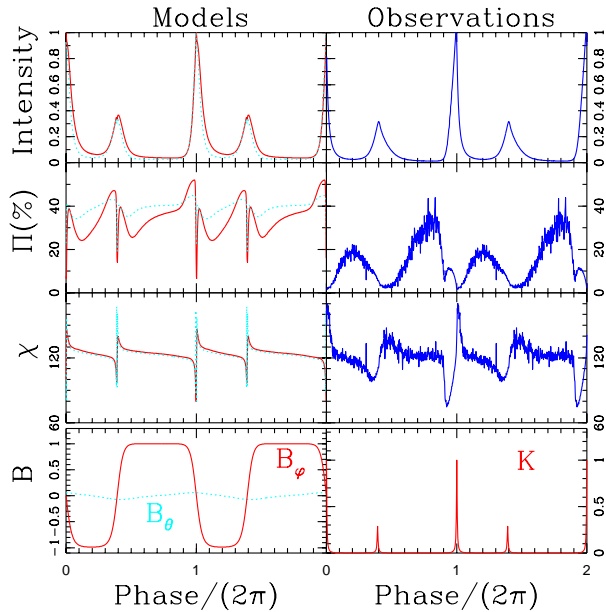}%
\caption{\footnotesize%
\label{polarisation}%
Model light curves of the intensity, degree of polarisation and 
  position angle of the pulsed synchrotron emission 
compared to measurements of the Crab pulsar \citep{kellner02,kanbachetal03}
(from \citet{petrikirk05}).
  Models with Lorentz factor~$\Gamma=20$ (solid line, red on-line) and
  $50$ (dotted line, cyan on-line) are shown.  The inclination of the line
  of sight equals the obliquity: $\alpha = \zeta = 60\degr$ and the
  position angle of the projection of the pulsar's rotation axis was
  set to $124\degr$ \citep{ngromani04}.  The bottom panels show
    the dependence on phase of the modelled magnetic field
    and particle density in the comoving frame.
}
\end{figure}

In the Crab pulsar, the  phase spacing of the 
the high-energy pulses is $0.4$ of a period. Combining this with the 
evidence, gathered from the nature
of the pulsed radio emission
\citep[see][]{kirkskjaeraasengallant02}, 
that the observer passes almost directly over
the magnetic pole, fixes both  
to be roughly $60\degr$. This agrees with an independent determination 
of $\zeta=61.3\pm0.1$ 
from the morphology of the X-ray torus \citep{ngromani04}. 

High time-resolution polarimetry of the Crab 
pulsar \citep{kellner02,kanbachetal03}  places additional, independent
constraints on this model. The high-energy pulsed 
emission (optical to gamma-ray) originates as the synchrotron radiation of
relativistic particles accelerated primarily in the current sheets. The 
magnetic field reverses direction across a sheet, but, in reality, this must
occur over a region of finite width. Within such a region, the density of
hot particles will rise, but the magnetic field intensity should
decrease. It may become turbulent, and, instead of a pure neutral sheet, any 
residual poloidal field component could cause a sweep of the field direction
rather than a discontinuous jump. To some extent, the 
population of heated particles
must also leak into the stripes surrounding the sheet,
in which case, the toroidal magnetic field 
component should dominate and manifest itself in the polarisation direction. 
These effects have recently been modelled \citep{petrikirk05} 
and compared to the data. The results, displayed in 
Fig.~\ref{polarisation} show that it is possible to obtain reasonably good
agreement for the intensity and degree of polarisation.  
For the angle of polarisation, very good agreement is achieved
with the observed sense and size of 
the polarisation sweeps through the pulses. 
Particularly noteworthy, however, is the excellent agreement
between the observed 
orientation of the linear polarisation vector between the pulses and
the predicted orientation.
Within the model, this 
is not sensitive to the choice of parameters, being 
coincident with
the projection of the rotation axis of the pulsar on the sky. This
is determined independently of the polarisation observations 
from measurements of the X-ray morphology of the nebula
\citep{ngromani04}. 

Thus, the \lq\lq smoking gun\rq\rq\ of magnetic energy dissipation in pulsar
winds may already have been detected \citep[see][]{arons04}. 
However, much work will
be needed to develop particle acceleration models in these sheets before it
becomes possible to estimate the calibre of the \lq\lq gun\rq\rq, where it
fires, and how the trigger mechanism works.  

\begin{figure}[t!]
\includegraphics[width=0.5\textwidth,clip=true]{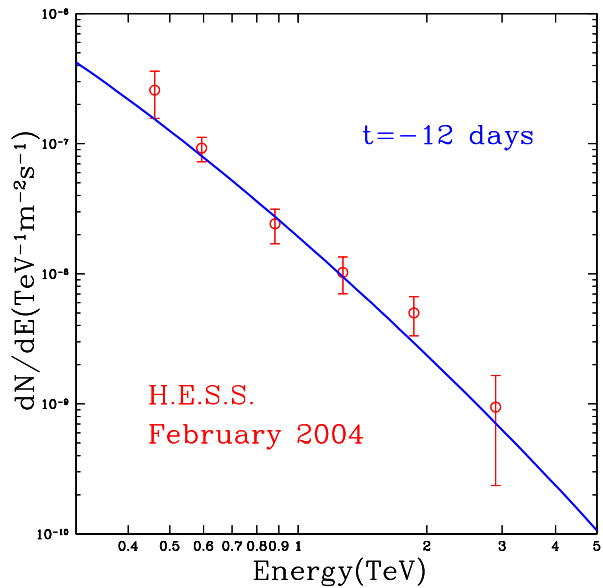}%
\caption{\footnotesize%
\label{1259}%
Predicted spectrum of inverse Compton scattered photons from the 
shocked pulsar wind plasma in the binary system PSR~1259-63, 12 days 
prior to periastron
\citep{kirkballskjaeraasen99}, and the 
spectrum observed by the H.E.S.S. array of imaging \v{C}erenkov detectors
in February 2004 \citep{schlenkeretal05,aharonianetal05}
}
\end{figure}
\section{High-energy emission from PSR~1259-63}

The prospect of additional observational constraints on the structure of a
pulsar wind has been opened up by the recent detection of high-energy (TeV)
gamma-rays from the Be-star/pulsar binary system PSR~1259-63 
by the H.E.S.S. array of imaging \v{C}erenkov telescopes 
\citep{schlenkeretal05,aharonianetal05}. This binary comprises 
a pulsar of 
period $48\,$ms in  an eccentric, 3.4 year orbit about 
a luminous Be star. Both stars presumably 
drive off winds, which interact, producing X-rays
\citep{tavaniarons97}. Furthermore, both winds originate on rapidly rotating
objects and are thought to have strong pole
to equator variations. Modelling of the unpulsed radio
emission suggests that the axis of rotation of the Be~star is not aligned with
that of the binary \citep{balletal99}, which complicates 
the dependence of the interaction on orbital phase. 
Relativistic electrons from the shocked pulsar wind are thought to be
responsible
for the X-rays, in which case they must also upscatter UV photons from the Be~star
to TeV energies. Modelling of this process led to the prediction
shown in Fig.~\ref{1259} for the multi-wavelength spectrum. The details
of the flow pattern of the emitting plasma are unknown, so that it cannot be
decided a priori, whether the radiating electrons suffer predominantly
radiative or adiabatic losses. However, provided the electron spectrum can be
\lq\lq calibrated\rq\rq\ using the 
synchrotron 
X-rays, this does not greatly influence the 
predicted TeV flux. 
The relative intensities of the
synchrotron and inverse Compton components, on the other hand, 
depend sensitively on the magnetic
field strength in the source. 
Fig.~\ref{1259} shows the predicted flux
for a magnetic field of $0.3\,$G, as indicated  
by models of the unpulsed radio emission
\citep{balletal99}, and in agreement with
estimates from the spin-down rate of the pulsar. The remarkable agreement in
both spectrum and normalisation indicates not only that the emission mechanism
was probably identified correctly, but also that the X-ray intensity and
magnetic field strength vary little from one periastron
passage to another.

In addition to the spectrum, the H.E.S.S. observations also revealed
substantial fluctuations in the light curve 
of this system
on a timescale of one day. These
cannot be accounted for within a model such as that of 
\citet{kirkballskjaeraasen99}, which 
assumes spherically symmetric stellar winds. However, 
fluctuations are certainly to be 
expected if either or both winds 
is asymmetric about the normal to the plane of the orbit, 
especially close to periastron, where the true anomaly varies 
rapidly. A correlation with the unpulsed radio flux \citep{johnstonetal05} 
would be important in this connection, possibly indicating a connection with 
the passage of the pulsar through the equatorial outflow zone of the 
Be~star \citep{kawachietal04}.  

The model underlying the prediction in Fig.~\ref{1259} assumes the 
Lorenz factor of the pulsar wind just before it enters the termination shock 
located between the two stars is a few times $10^7$. 
At periastron, this point is, at most, a few thousand light 
cylinder radii from the pulsar, and can grow to a few hundred thousand at
apastron --- minute compared to the $10^9\rlight$ stand-off distance of the
termination shock of the Crab pulsar. 
At first sight, therefore, the relatively high Lorentz factor 
appears to 
favour more rapid conversion of magnetic energy to particle energy than could 
be accommodated in the gradual acceleration models of 
\cite{kirkskjaeraasen03}. But the extent to which it will be possible to 
test pulsar wind acceleration theories, for example by placing firm 
limitations on the radial variation of the Lorentz factor, will depend 
on the availability of simultaneous X-ray and hard gamma-ray observations 
throughout the orbit, as well as the development of more sophisticated 
spectral modelling techniques.  
\begin{acknowledgements}
I thank J.~Arons, J.~P\`etri and Y.~Lyubarsky for helpful discussions.
This research was supported in part by the National Science Foundation under
Grant No.~PHY99-0794 and by a grant from the G.I.F., the German-Israeli
Foundation for Scientific Research and Development.
\end{acknowledgements}

\end{document}